\documentstyle[epsf,onecolumn]{mn}
\newcommand{\be}{\begin{equation}}
\newcommand{\ee}{\end{equation}}
\def\de{\hbox{d}}
\newcommand{\spose}[1]{\hbox to 0pt{#1\hss}}
\newcommand{\lta}[0]{{\mathrel{\spose{\lower 3pt\hbox{$\mathchar"218$}}
     \raise 2.0pt\hbox{$\mathchar"13C$}}}}
\newcommand{\gta}[0]{{\mathrel{\spose{\lower 3pt\hbox{$\mathchar"218$}}
     \raise 2.0pt\hbox{$\mathchar"13E$}}}}

\begin{document}
\title{On the Evolution of the Galactic Globular Cluster System}
\author[E.Vesperini]{E.Vesperini\thanks{Present address: Department of
Physics and Astronomy, University of Massachusetts, Amherst, MA 01003-4525, USA
}\\
 Scuola Normale Superiore  Piazza dei Cavalieri 7, 56126-I, Pisa\\
E-mail:vesperin@astro.sns.it
}
\maketitle
\begin{abstract}
By means of a semi-analytical method allowing us to follow the
evolution of individual globular clusters spanning a large set of
different initial conditions we address the issue of the origin of
some observational properties of the galactic globular cluster system.

After a preliminary study of some general properties of the main
evolutionary processes by a discussion of the relevant timescales and
of a suitable ``survival factor'', we investigate the evolution of
systems of globular clusters located in 
a  model of the Milky Way  starting from different initial
conditions for the mass function of the system (power-law and log-normal) and
for the  distribution of concentrations of individual clusters.
 
In particular, we study  the role of the
evolutionary processes in changing the spatial distribution and mass function
of the cluster system, in establishing  and/or preserving some of the observed
correlations and trends between internal properties of globular clusters and 
between internal properties and location inside the host galaxy and we provide
 an estimate for the rates of core collapse and disruption of globular
clusters. 

For the initial conditions considered in this paper a significant fraction  of
clusters ($\sim 60\%$) is lost because of disruption; the fraction of those
undergoing  core collapse is consistent with the present observational limits.
The initial mass function and spatial distribution of the cluster
system evolve quite significantly in one Hubble time and the evolution
is toward a final state  similar
to the observed one. If the  mass 
function is initially taken to be a log-normal distribution similar to the
 one currently observed in our galaxy, its shape is not significantly altered
during the entire simulation  even though a significant number of clusters are
disrupted before one Hubble time, which suggests that the present mass function
might represent a sort of 'quasi-equilibrium' distribution.
 \end{abstract}   



\parindent=1pc

\section{Introduction}
Recent surveys of the observational properties of the globular clusters of our
Galaxy (see, e.g., Djorgovski \& Meylan 1994, hereafter DM) have shown the
existence of interesting correlations and trends between structural parameters
and between structural parameters and location inside the Galaxy, suggesting a
close connection  between dynamical evolution of clusters and the galactic
environment of the host galaxy.

 Many of the observed correlations still lack
 a solid theoretical interpretation; in particular it remains unclear  to what
extent the present properties of the galactic globular cluster system reflect the
initial conditions and to what extent they have been determined by evolutionary
processes causing the complete evaporation of a   fraction of primordial
clusters and the modification of the initial properties of the surviving ones.
While many theoretical studies suggest that the present cluster system, as the
relic of the initial system (see, e.g., Aguilar et al. 1988, Lee \& Goodman 1995,
van den Bergh 1993, Murali \& Weinberg 1996, Gnedin \& Ostriker 1997),
 would have properties 
substantially different from those determined by formation processes, some
observational evidence, such as the similarity of the luminosity function in
galaxies of different types and different sizes (see, e.g., Harris 1991), is an
indication that the local environment might be not so important.

In this paper we provide a simple theoretical framework for the investigation of
 the evolution of
globular clusters. Here we extend and improve some of the tools and arguments
considered in previous articles that addressed the issue of  the efficiency of
evolutionary processes in terms of an  analysis of the relevant timescales  (Fall
\& Rees 1977, Caputo \& Castellani 1984) or on the same method adopted here
(Chernoff, Kochanek \& Shapiro 1986, hereafter CKS, Chernoff \& Shapiro 1987) 
and show the results of a set of simulations following the
evolution of the properties of a globular cluster system (mass
function, spatial distribution, correlations between structural
parameters) starting  from  given 
initial conditions. The main goal of the analysis is that of establishing the
role of initial conditions and  evolutionary processes in determining the present
observational properties.

 In section 2  we  set the empirical background for our theoretical analysis.
 In section 3 we describe the method  adopted to follow
the evolution of individual globular clusters under  the effect of internal
relaxation and disk shocking. In section 4
we discuss the behaviour of the time scales associated with the above processes
and introduce a survival factor describing the regions of the Galaxy where
clusters of a given mass are more likely to survive and those where they are more
efficiently destroyed. In section 5 we show the main results of some
simulations which we have  carried out by the method described in
section 3 and by which we investigate the evolution of the main
properties of globular 
cluster systems starting from specific initial conditions.\\
Summary and conclusions are in section 6.
 \section{Empirical background and goals of the
theoretical investigation}
The results of several observational studies of the Galactic globular
cluster system (Chernoff \& Djorgovski 1989, Djorgovski 1991, Trager,
Djorgovski \& King 1993,  DM, Bellazzini et al. 1996 and references therein)
can be summarized in the following way:
 \begin{itemize} \item Core parameters vs. total luminosity\\
More luminous
clusters have higher concentrations (Djorgovski 1991, DM, van den Bergh 1994),
smaller cores (DM) and higher central densities (DM); the large spread
present suggests the existence of a second parameter in the above correlations.
As to the trend between core radius and luminosity, it is interesting to note
that it is opposite to that observed for elliptical galaxies for which the size
of the core increases with the  luminosity of the system (see, e.g., Lauer 1985).

 It has been suggested
(Djorgovski 1991) that  the correlation concentration-luminosity   might be
explained in terms of initial conditions for what concerns the lack of
low-concentration high-luminosity clusters, while evolutionary processes should
account for the lack of high concentration low-luminosity clusters.
In DM the correlation has been
interpreted "..as a differential survival effect, with more massive clusters
surviving longer and reaching more evolved dynamical states". 

On the other hand  the same
correlation is present in a sample of dwarf
ellipticals (Binggeli, Sandage \& Tarenghi 1984 figure 10b, Ichikawa, Wakamatsu
\& Okamura 1986 figure 11) for which evolutionary processes should be less
important  and,
 as shown by Bellazzini et al. (1996), the correlation is stronger
for globular clusters located at larger distances from the galactic center where
evolutionary processes act on longer timescales (see section 4).

\item Spatial distribution.\\
The number density of clusters, $n$ (number of clusters per cubic kiloparsec), is
well fitted for galactocentric distances $R_g>3$ kpc by a power-law $n(R_g)
\propto R_g^{-3.5}$ (see, e.g., figure 2 in Zinn 1985). The distribution shows a
flattening in the inner regions whose origin is uncertain: the fact that
clusters appear to be missing may be due to obscuration, to an
intrinsic flattening of the distribution or to the effects of tidal destruction.

\item Trends of structural parameters with the position in the Galaxy.\\
Clusters closer to the galactic center tend to have higher concentrations and
smaller cores (see figure 9 in DM). Moreover Chernoff \& Djorgovski (1989)
(see their figure 9) show that the fraction of post-core-collapse (PCC)
clusters is a strong function of the distance from the galactic center with 
most  PCC clusters  located in the inner regions of the Galaxy. 
\item
Luminosity function.\\ The luminosity function of  the Galactic globular cluster
system is well fitted by a Gaussian with peak value $\langle M_V \rangle
=-7.24\pm 0.13$ and variance $\sigma=1.1$ (Secker 1992; see also Abraham \& van
den Bergh 1995).
Harris \& Pudritz (1994) have shown that the luminosity function can  be
adequately described also by a multiple index power-law in the luminosity $L$ and
have suggested a possible origin for such a shape assuming that clusters have
been originated from the cores of supergiant molecular clouds present in the
early protogalactic epoch.
\end{itemize}
 
In this work
we will try to clarify {\it which of the observed correlations and
properties of the galactic globular cluster system are likely to be
the relic of  primordial initial conditions and which instead are likely to be 
essentially independent of the detailed initial conditions and must
have been mostly determined by evolutionary processes}. 

Special attention will be paid to:

-	Characterization of the most important internal and 
external processes (with respect to a given cluster) that 
determine dynamical evolution and development of simple and 
flexible tools for their description, to be incorporated in a study 
of the evolution of a system of globular clusters from given 
initial conditions.

-	Efficiency of the various evolutionary processes in 
affecting the results of given initial conditions and in erasing the 
memory of such conditions.

-	Evolution of the shape and of the parameters that 
characterize the luminosity function of the globular cluster 
system.

-	General correspondence with some observed correlations 
and trends.

\section{Evolving a system of globular clusters from given initial conditions}
\subsection{Evolution of an individual  cluster}
The general method followed in the present investigation, which was  introduced
by King (1966) and later  used by Prata (1971a,b), CKS and
Chernoff \& Shapiro (1987), is based on the assumption that the evolution  of
individual globular clusters can be approximately described as a sequence of
King models with different concentrations, masses and radii.

 A King model (King 1965) is
identified by three parameters, the total energy $E_{tot}$, the total mass $M$
and the truncation radius $r_t$ which we shall assume to be determined by the
tidal field of the galaxy (see, e.g., Spitzer 1987) and defined as  
\be
r_t=\left(M\over 3M_g\right)^{1/3}R_g \label{tid},
\ee
where $R_g$ is the radius of the orbit of the globular cluster around the host
galaxy and $M_g$ is the mass of the galaxy contained inside that radius.
For simplicity, the cluster orbits are taken to be circular.
After a given time interval, the change of mass and energy  produced by
evolutionary processes can be calculated and a new tidal radius can then be
determined by the tidal condition (\ref{tid}).
The total change in mass and energy is given by the sum of the variations of
these quantities due to all the processes included in the investigation; 
in our investigation we have included the effects of disk shocking and
internal relaxation. 

 The three new values of mass,
energy and radius identify a new King model with different
concentration and different scale parameters. The evolution of each cluster
is followed until one of these three conditions is satisfied:
 \begin{itemize}
\item  $W_0 > 7.4$ ($W_0$ indicates the dimensionless central potential of
a King model). Both analytical
studies and numerical integrations of the Fokker-Planck equation (Katz 1980,
Wiyanto, Kato \& Inagaki 1985) indicate that King models with $W_0$ larger than
this value are unstable against gravothermal catastrophe. Gravothermal collapse
is driven by a thermodynamic instability and the description of the  evolution
of the cluster parameters during this phase is outside the reach of this
method. Moreover the structure of a cluster in the PCC
phase and its evolution are well known to depend significantly on the
source of energy supporting the PCC expansion (3-body, tidal or
primordial binaries), and in the case in which primordial binaries are
the relevant source on the fraction and the binding energy
distribution of these (Heggie \& Aarseth 1992, Vesperini \& Chernoff
1994). A reliable modelling of these aspects, as well as of other
processes (e.g. gravothermal oscillations), is beyond the
possibilities of the method adopted for our investigation.

\item $W_0<0.05$. Under this value the system is  conventionally considered
dissolved (Chernoff \& Shapiro 1987).

\item $t=1.5 \times 10^{10}$ years (Hubble time). The present era has been
reached.
\end{itemize}

In the simulations discussed in section 5 we have also taken into account the
effects of dynamical friction, even though in a very approximate way, by
eliminating from the final sample of survived clusters all those for which the
orbital decay timescale (see, e.g., Binney \& Tremaine 1987) is
smaller than one 
Hubble time. Given the initial conditions chosen  the inclusion of dynamical
friction does not change significantly the final results.

As a test for the
numerical code developed in our work, some runs with the evaporation rate used
in CKS have been carried out and the time evolution of the concentration shown
in their figure [11a-c] has been reproduced. 

It is  clear that our final sample of globular clusters will
include  only those clusters that can still be represented as King models after
one Hubble time, while the method cannot predict anything about the state of
those which underwent gravothermal catastrophe  before that time.

The main limitations of our approach are the impossibility of following the
evolution of clusters after the onset of gravothermal catastrophe and the
 fact that we have restricted our study to  circular orbits. The latter
point suggests that some caution should be used when comparison is made with
the observations, since some of the present results on the trends with
galactocentric distance are expected to be affected by orbit mixing. On the
other hand, our approach has the advantage of allowing us to single out certain
dynamical effects in a model where only a few parameters are varied. The
evolution of clusters with a more general distribution of orbits will be
considered in a future work but, of course, at the cost of increasing
significantly the relevant parameter space of initial conditions.

 \subsection{Internal
relaxation}
 As a result of repeated weak encounters, stars can
gain  sufficient energy to escape from the system. Due to the diffusive nature of
this process the energy  of a single star escaping from the system is
equal just to that necessary to reach the tidal radius where the cluster ends.
This means that escaping stars do not carry away kinetic energy and that the
change in cluster energy due to the evaporation of stars from the cluster driven
by internal relaxation is   
\be
\Delta E_{tot}=-{GM \Delta M \over r_t}\label{erel}.
\ee

An estimate of the evaporation rate can be made using the Fokker-Planck
equation. The evaporation rate is calculated as the flux of particles
through the boundary in the energy space, $E_t$
\be
{\partial N \over \partial t}= -16 \pi^2\left( 4 \pi \Gamma {\partial f \over
\partial E}\vert_{E_t} \int_{E_0}^{E_t}f q \hbox{d} E \right),
\ee
where $\Gamma \equiv 4 \pi G^2m^2\ln(N/2)$;  $q \equiv {1
\over 3} \int_0^{r_{max}}[2(E-\Phi)]^{3/2}r^2 \hbox{d} r$ ($\Phi(r)$ is
the potential energy of a star at a radius $r$ and $r_{max}$ is the radius where
$\Phi(r_{max})=E$ and we take $f$ equal to the King distribution function.
This can be written in the form
\be
{\partial N \over \partial t}=- ( 16 \pi^2)^2 \sqrt{G} \ln (N/2)
\sqrt{M/r_t^3}F( W_0)
\ee
where $F(W_0)$  depends only on the central dimensionless potential $W_0$.
The
mass loss due to internal relaxation during a lapse of time $\Delta t$ is
\be
\Delta M=\langle m \rangle {\partial N \over \partial t} \Delta t.\label{mrel}
\ee
In the following we will generally refer to  $\langle m\rangle =0.7 M_{\odot}$.
The evaporation rate derived above is a factor 2-3 (depending on $W_0$)
smaller than that used in CKS following King (1966) which overestimates the real
rate as it is correct only in the central regions of the cluster.

We define the characteristic timescale for evaporation
\be
T_{ev} \equiv \vert{N \over \partial N /\partial t}\vert= {2.27 \times 10^{12}
\hbox{yrs} \over \ln(N/2)\left(\langle m \rangle \over 0.7 M_{\odot}\right)}\left( {M \over 10^5
M_{\odot}}\right)^{1/2} \left( {r_t \over 50 \hbox{pc}}\right)^{3/2}
\hat{T}_{ev}(W_0) 
\ee
where $\hat{T}_{ev}(W_0) $  depends only on the central dimensionless potential
$W_0$ and it covers a range of values between 0.2 and 0.35.

Replacing $r_t$ by  its expression given by the tidal boundary condition
(eq.(\ref{tid})) and assuming an 'isothermal' model for the host galaxy
($M_g/R_g=const$) we have
\be
T_{ev} \propto MR_g \hat{T}_{ev}(W_0) .
\ee
 \subsection{ Model for the host galaxy and disk shocking}
When a globular cluster, during its orbital motion around the host galaxy,
crosses the disk, due to the tidal interaction  the stars, on average, increase
their kinetic energy and some of them gain energy enough to escape from the
cluster. If we select the direction of the  $z$ axis as that orthogonal to
the galactic disk and approximate the motion of a single star inside the cluster
in this direction as that of a harmonic oscillator with frequency $\omega$ we can
write the equation of the motion for the single star along the $z$ axis,
including the perturbation due to the presence of the disk, as
\be
{\hbox{d}^2 \delta z \over \de t^2}+\omega^2 \delta z=g(t)\delta z
\ee
where $g(t)=\left({\de K_z(R,z)\over \de z}\right)$, $K_z(R,z)$ being the
  acceleration due to the gravitational field of the disk  at a
given galactocentric distance $R$ and at a distance from the disk of the
center of the globular cluster $z=Vt \cos \Theta$  (here $t=0$ denotes the
instant when the cluster crosses the equatorial plane of the disk), $\Theta$ is
the angle between the direction of motion of the cluster and the $z$ axis, $V$
is the orbital velocity of the globular cluster, taken to be constant $V \simeq
210 \hbox{km s}^{-1}$ corresponding to a constant ratio $M_g/R_g \simeq 10^7
\hbox{M}_{\odot}/\hbox{pc}$ and $\delta z$ is the distance of the star from the
center of the globular cluster along the $z$ axis.

 For
the potential of  the disk of our Galaxy we have used the one obtained by
CKS by fitting with a two-component 'isothermal' model the
acceleration along the $z$ direction in the solar neighbourhood 
 determined by Bahcall (1984) (taken to be at
a galactocentric distance $R_0$ equal to 8 kpc).  

The vertical acceleration, $K_z$, varies with the galactocentric
distance according to the surface 
density that, according to the Bahcall-Schmidt-Soneira (1982) model for the
Galaxy we will adopt in our investigation, falls off exponentially with an
exponential scalelength $h=3.5$ kpc. 
The change of kinetic energy
$\Delta e$ per unit mass of a single star produced by the interaction with the
potential of the disk can be shown to be 
\be
\Delta e={\langle\Delta v_z^2\rangle \over 2}={\pi^2 \over 8}
\langle \delta z^2\rangle \omega^2\left[{K_0 \over z_0 \nu_0^2} {1\over \sinh(\pi
\omega/2 \nu_0)}+{K_1 \over z_1 \nu_1^2} {1\over \sinh(\pi
\omega/2 \nu_1)}\right]^2 \left({e^{-R_g/h} \over e^{-R_0/h}} \right)^2
\ee
where  $K_0=3.47 \times 10^{-9} \hbox{cm s}^{-2}$, $z_0=175~$pc,
$K_1=3.96 \times 10^{-9} \hbox{cm s}^{-2}$, $z_1=550~$pc are the parameters for
the disk model obtained by CKS and $\nu_0={V\cos \Theta
\over 2 z_0}$ and $\nu_1={V\cos \Theta \over 2 z_1}$  are the characteristic
frequencies of disk crossing by the globular cluster. For stars with
$\omega\ll\nu$ (impulsive limit) $\Delta e$ is maximum, while if $\omega\gg\nu$
(adiabatic limit) $\Delta e$ tends to zero.
 Assuming
\be
\langle  \delta z^2\rangle ={r^2 \over 3}
\ee
and 
\be
\omega^2={GM(r) \over r^3}
\ee
we get that $\Delta e$ depends only on the distance from the centre of the
cluster.

The input of kinetic energy in the cluster due to the interaction with the disk
is thus calculated as 
\be
\Delta T= 4 \pi \int_0^{r_t}\rho(r) \Delta e(r)r^2 \de r.
\ee
The integral includes only those stars that do not evaporate from the cluster
after the interaction with the disk.
The total change in energy is
\be
\Delta E_{ds}=-{GM \Delta M_{ds} \over r_t}+ \Delta T \label{edi}.
\ee
where $\Delta M_{ds}$ is the mass loss due to disk shocking. For the process of
disk shocking we can define a characteristic time scale as 
\be 
T_{ds}=\vert{E \over
\partial E /\partial t}\vert \propto \sqrt{{R_g^3\over
GM_g}}\hat{T}_{ds}(W_0,M,r_t, \omega/\nu_i,K_i,R_g,h) 
\ee
where $\partial E /\partial t$ is calculated as the ratio of the change of
energy during one interaction with the disk (eq. \ref{edi}) to half of the
orbital period $T_{orb}/2=\pi\sqrt{{R_g^3\over
GM_g}}$ and in the function $\hat{T}_{ds}$ we have
included both the dependence on the internal structure of the cluster and the
structure of the disk. Fixing the structure parameters of the disk and using the
tidal boundary relation (\ref{tid}), we find that the disk shocking timescale
depends only on the galactocentric distance and cluster concentration 
\be 
T_{ds} \propto \tilde{T}_{ds}(R_g,W_0) 
\ee

\section{Time scales and survival factor in the natural parameter space}

The two timescales, $T_{ev}$ and $T_{ds}$, introduced in the previous
sections are plotted in 
Figure 1 as functions of the galactocentric distance.

The timescales plotted represent the average of values corresponding to 
different values
$W_0$ ($W_0=1,~3,~4,~7$). For what concerns the timescale associated with
internal relaxation,  values corresponding to $M=10^4 M_{\odot}$, $M=10^5
M_{\odot}$ and $M=10^6 M_{\odot}$ have been plotted. It is clear that disk
shocking is more important for more massive clusters and particularly at
distances close to $R_g \sim 4$ kpc where disk shocking reaches its maximum
efficiency. For galactocentric distances smaller than 4 kpc, as $\omega$
increases to a larger extent than $\nu$, the heating from disk shocking is
adiabatically suppressed and the disk shocking timescale increases. 

The conclusion of CKS and Chernoff \& Shapiro (1987) concerning the
faster dynamical evolution of clusters near the galactic center is
confirmed by the behaviour of timescales. It is also evident that   in
the inner regions of 
the Galaxy, survival of less massive clusters is more difficult and
 we can expect that, as a result of dynamical evolution, massive
clusters might become dominant in the inner regions. In the outer regions, all
timescales are greater than the Hubble time and the processes driving the
dynamical evolution of globular clusters   become less and less
efficient and we 
can expect that properties of the globular cluster system will be closer to the
primordial conditions.

 These qualitative expectations depend, of course, on the
initial mass function of the globular cluster system: an initial mass function
including many very massive clusters would be unaltered also in the inner
regions  (except for the effects of dynamical friction), while a 'weak' initial
mass function including many low-mass clusters would be strongly affected also
in the outer regions.

In order to make the above considerations  more quantitative we have
followed the evolution of clusters at a galactocentric distance between 1 kpc and
20 kpc starting with $W_0$ in the range $0.6<W_0<7.25$  and defined a
{\it survival factor}, $\xi$, as the ratio of the number of clusters that after a
Hubble time have not yet undergone core collapse or dissolution to the total
number of initial conditions considered. For this general calculation we have
assumed that all the values of $W_0$ are equiprobable. 

The survival parameter depends on the mass of the cluster and on the
galactocentric distance. For each value of the mass of the globular cluster
considered we have investigated the dependence of the survival factor on the
galactocentric distance. In Figure 2 the survival factor versus $R_g$ is plotted
for six different values of  $M$. For less massive clusters $\xi$ is
different from zero only for the higher values of $R_g$. For high-mass clusters
disk shocking becomes  important and the behaviour of $\xi$
resembles that of disk shocking timescale with a minimum close to $R_g \sim 4$
kpc where disk shocking has its maximum efficiency.

 We emphasize that the
survival factor $\xi$ we have calculated is intended to give a general
quantitative indication of the trends induced by dynamical evolution.
The
behaviour of the survival factor and of the timescales confirms the
conclusion drawn by CKS about the larger efficiency of evolutionary
processes in the inner regions of the Galaxy. Inner clusters are
destroyed more easily and this might account for at least some of the observed
flattening in the  number density of clusters 
and for the trend for more evolved clusters to be located closer to the galactic
center; moreover less massive clusters are destroyed in the
inner regions to a larger extent than the more massive ones and a
trend for clusters in the inner regions of the Galaxy to be on the
average more massive than outer clusters might result from the action
of evolutionary processes.

Further interesting conclusions can be drawn from  Figures 3-4 providing
additional information on the fate of clusters of different initial masses
located at different galactocentric distances. Evolution of clusters starting
with different values of $W_0$ and $R_g$ has been followed: Figure 3 shows
only those values of the initial conditions such that clusters survive
disruption and do not undergo  core collapse.  An
analogous plot has been recently shown by Weinberg (1994a) (see his figure
9)  who  has implemented a Fokker-Planck code including accurate estimates of
the effects of disk shocking which, moreover, have been modelled by
means of a new expression of the adiabatic correction derived in
Weinberg (1994b).
While Weinberg's numerical model is able to follow the evolution of individual
globular clusters in larger detail (see also Murali \& Weinberg 1996)
than we can do with the method adopted in 
this work, our method has the advantage  of requiring a very small
computer time for each model investigated thus allowing the investigation of a
much larger set of different initial conditions; the survival region (to
disruption) as obtained by Weinberg, for an initial mass of the clusters $M=3
\times 10^5 M_{\odot}$  essentially coincides with the one obtained in our work
for the same mass lending support to the reliability of our calculations.

Figure 4 shows the final values of $W_0$ versus $R_g$ for the survived
clusters. It is interesting to note that if more massive
clusters
 with low values of $W_0$ (low concentrations) were initially present, they are
expected to survive with such low values of $W_0$ at larger galactocentric
distances; some low-concentration massive clusters are also expected to be
present in the inner regions due to the effects of disk shocking driving the
clusters toward the dissolution. 
It is important to note also that if
some high-concentration, low-mass clusters are present at the beginning they
will still be present in the present epoch.

Thus the results summarized in Figure 4 lend support to the hypothesis that
{\it the observed correlation between concentration and mass is primordial and
cannot be the result of evolutionary processes}.

\section{Evolution of a system of globular clusters}
\subsection{Choice of Initial conditions}

 The present
knowledge of the formation of globular clusters (see, e.g., Fall \& Rees 1985,
Vietri \& Pesce 1995) and of the detailed initial conditions of the
galactic globular cluster system is rather poor and it is not clear whether
the present globular cluster system keeps some trace of its initial
properties or it has largely lost
memory of the initial conditions  as a result of evolutionary processes. On the
other hand,  the analysis of the characteristic time scales associated with the
evolutionary processes shows that evolutionary effects are
likely to be small for clusters located in the outer parts of the Galaxy. Thus a
promising procedure could be to assume that the initial properties of globular
clusters do not vary with galactocentric distance to get from the current
properties of clusters at large galactocentric distances  some indication
on the initial conditions for the entire system.

 For each run carried out in
our investigation the evolution of 1000 clusters has been followed by the method
described in sect. 3. The set of initial conditions has been produced  according
to the  distributions described below in this section and the model
adopted for the Milky Way is that described in section 3.

Whenever we  compare our results with the observational data, we will refer to
the data of Chernoff \& Djorgovski (1989) restricted to clusters with $R_g<20$
kpc and $c\lta 1.7$ as for clusters in our theoretical analysis. The mass of
clusters is calculated from their luminosity based on  a constant mass-to
light ratio $M/L_V=2$ (see, e.g., Mandushev, Spassova \& Staneva 1991).

A globular cluster system is characterized once the following
properties  are specified:
\begin{itemize}

\item Initial mass function.\\ 
Three different initial mass functions have been considered:
\begin{itemize}

\item A truncated power-law mass function $f(M) \propto M^{-2}$ with lower
cut-off $M_{low}=10^{4.5} M_{\odot}$ and upper cut-off $M_{up}=10^{6} M_{\odot}$
(POW). 

\item A Gaussian distribution in the logarithm of the mass with mean value $<\log
M>=4.5$ and variance $\sigma^2=1.0$ (GAU1).

\item A Gaussian distribution in the logarithm of the mass with mean value $<\log
M>=5.0$ and variance $\sigma^2=0.25$ (GAU2).
\end{itemize}

 The motivation  for the choices made in our investigation can be
summarized by the following  questions:

1)  Given a  power-law
initial mass function how does evolution change it? Can evolution drive it to a
multiple-index power law or is this shape the result of formation processes
as suggested by Harris \& Pudritz (1994) ?

 2) Given a  Gaussian initial mass function, will this shape be
preserved during evolution? How does its mean value  evolve and in which
direction?

3) How much do the final mass functions in the three cases differ from each
other? 

As for the GAU2 case, we point out that the mass function of this case
is very similar 
to the present mass function: the reason for investigating this case is that we
wonder whether it is plausible that the initial mass function was essentially
similar to the present one and  to what extent it was modified by
evolutionary processes. 

We stress that, given the very
limited number of different initial mass functions investigated, our
analysis is not meant to provide any definitive indication on the
entire possible set of initial conditions eventually evolving into a
state resembling that currently observed for the galactic globular
cluster system; a more extensive investigation spanning a larger set
of different initial values of the parameters of the initial mass
function, aimed to draw more specific conclusions on this issue, is
currently in progress (Vesperini 1996, in preparation).

Of course, besides the above points, it will be interesting to investigate how
all the final results, involving structural parameters, depend on the initial
mass function.

\item Spatial distribution.\\ 
The number density profile is chosen to be $n(R_g)
\propto R_g^{-3.5}$ according to what is observed in our Galaxy  between
$4<R_g<20$ kpc, where the sample of observed globular clusters is likely to be
complete (Zinn 1985).

\item Distribution of orbital parameters.\\
For simplicity we have limited our analysis to
circular orbits crossing the plane of the galactic disk perpendicularly.

\item  Initial distribution of $c=\log r_t/r_c$.

An interesting feature of the galactic globular cluster system
is the correlation $c-\log M$.  
The results discussed in sect. 4 (see also Bellazzini et al. 1996)
show  that evolutionary  processes are unlikely to play a dominant role in
establishing the $c-\log M$ correlation. Thus  we have decided to
assume among the initial conditions  a relation between $c$ and
$\log M$. The relation sets an initial trend on the basis of
the properties of the more massive and distant clusters  (probably less affected
by evolutionary processes)   
\be  
 c=-2+0.6 \log M
\ee
Notice that it is not obvious whether such a correlation  would be preserved
during  evolution or not.
In order to test the hypothesis that evolutionary processes might play some
role in establishing the observed correlation (Djorgovski 1991, DM), we have
also carried out  two simulations starting from an initial condition with 
uncorrelated $c$ and $ \log M$ (POW-nc and GAU1-nc).   
\end{itemize} 
 
\subsection{Results}
\subsubsection{ Correlations between structural parameters} 
 Figure 5 shows the final $c-\log M$ plane for the POW run (similar results
have been obtained for the GAU1 and the GAU2 runs) with  the line showing the
initial condition of the simulation. It is
interesting to point out that the evolutionary processes give rise to a spread
similar to that present in the observational data. It is evident that the effect
of evolutionary processes is that of scattering initial conditions: clusters
undergo different evolution in the $c-\log M$ plane toward larger or smaller
concentration and lose a different fraction of their initial mass  depending on
the   efficiency of disk shocking and internal relaxation.
In agreement with the trend emerging from  the analysis of observational data
(Bellazzini et al. 1996), in all three simulations, the correlation  $c-\log M$
is stronger for outer clusters which preserve better the memory of the initial
conditions while for those closer to the galactic center evolution is faster and
tends to smear more efficiently the initial correlation. Our results give
strong support to
 the hypothesis that the correlation between concentration and mass of
clusters, or more in general between core parameters and mass (see Bellazzini et
al. 1996), is primordial and not induced by evolutionary processes.   

The results of the runs GAU1-nc and POW-nc strengthen  this view further:
Figure 6  (from the GAU1-nc run;
an analogous result is obtained in the POW-nc run) shows that, if $c$ and $\log
M$ are initially uncorrelated, also at the end of the simulation there is
neither a correlation between these two quantities nor even a trend to create
the correlation.\\

\subsubsection{Mass function}
Figures 7a-c show the initial and the final  histogram of the
globular clusters  mass function  for the three  runs, while Figures 8a-c
show the initial, the final and the observational cumulative distribution
function (CDF) of $\log M$. 
It is worth noting that all the final samples are
just the relic of larger  initial systems. Of course  a 'stronger' initial mass
function, including a larger number of high mass globular clusters, would be
largely unaffected by evolution and its final state would be much closer to the
primeval conditions.

Evolution changes the power-law
initial mass function into a bell shaped mass function, resembling a Gaussian
mass function in $\log M$. 
Moreover from Figure 9 it is evident that evolution modifies the
initial power-law distribution in a two-component power law distribution in $M$
consistent with the analysis of
Harris \& Pudritz (1994; see also references therein) who show that the mass
function of galactic globular clusters is almost flat in the low-mass tail and a
power-law with index $n=-1.7$ for masses $M>0.75 \times 10^5 M_{\odot}$. 
 The observed modification of the
power-law mass function toward multiple power-law is  equivalent to  the
trend for the distribution to assume a bell-shaped form resembling a Gaussian in
$\log M$ (see also McLaughlin 1994). 

In the GAU1 run the gaussian shape is essentially preserved, 
but shifted to a higher mean value. The low-mass side is almost entirely
destroyed by the evolutionary processes. Even though a thorough comparison of
our final results with observational data is beyond the scope of this analysis,
 nevertheless  
the agreement between the final CDF for the GAU1 run and the observational
one is remarkable.   
The similarity between the two distributions is even more remarkable if one
considers how different  the initial distribution is from the observed one.

The low-mass
tail obtained in the final mass function of the POW run is produced by loss of
mass of the less massive clusters present in the initial mass function while
for the GAU1 run the low-mass tail is what is left of  the much larger low-mass
side of the initial mass function.

The result of the GAU2 is extremely intriguing. It is evident  from
Figures 7c and 8c that, though about $70 \%$ of the initial  number of clusters
undergo disruption or core collapse,  the   shape of the mass function is
essentially preserved during the entire evolution. The result of this simulation
shows the surprising result that  {\it a significant effect of evolutionary
processes does not necessary imply a strong difference between initial and final
mass function}.  It is tempting to suggest that the present mass function
represents a sort of "dynamical equilibrium"; this hypothesis seems to
be further 
supported by the following points:
 \begin{itemize}
\item Figure 10a shows the time evolution of the mean value of $\log M$
for the runs GAU1 and GAU2. $\langle \log M \rangle$ for the GAU2 run is
approximately constant during the entire evolution, even though, as 
pointed out earlier, a significant number of clusters undergoes disruption or
core collapse and thus are being excluded from the sample;  $\langle \log M
\rangle$ for the GAU1 undergoes a significant change until $\sim 4-5$ Gyr when
it reaches a value close to that of the GAU2 run and  after that it is almost
constant and equal to  the 'equilibrium' value. Figure 10b shows the time
evolution of the total number of clusters.
\item Figure 11 shows the
CDF of $\log M$ for the GAU1 run at five different times $t=0,~2,~4,~9,~15$
Gyr; it is evident that after an initial lapse of time during which the
distribution undergoes a significant change, it reaches  a quasi-equilibrium
state similar to the observational distribution  and to the initial and final
distribution of the GAU2 run.  
\end{itemize}  

The rate of evolution of the mass function of the globular cluster system is
determined in part by the properties of the mass function itself and in part by
the underlying spatial distribution of the clusters.

The properties of a mass function can be in a state of approximate
``dynamic equilibrium'', if the number of disrupted clusters is suitably
compensated for by an appropriate mass
evolution of the surviving ones as it appears to occur in the GAU2
run. On the other 
hand, as a cluster system evolves, the destruction of clusters at small
galactocentric distances and of low-mass clusters in general tends to leave
a population of clusters more resistant to evolutionary processes and 
eventually a sort of equilibrium  can be achieved, which is approximately
``static", at least on the long Hubble timescale.

Obviously the mass function would also be in equilibrium on the Hubble
timescale  (a ``static
equilibrium" right from the beginning, in this case) if all or most of the
clusters were initially located at very large galactocentric
distances. In this case 
not only the shape of the mass function would be preserved but also the
total number of cluster would be essentially unchanged. Such equilibrium
would be a trivial result expected for any initial condition. What we find
in the GAU2 run is, instead, a major surprise since the mass function is
found to persist not because of a particularly favourable underlying
spatial distribution, but as a results of a subtle balance of competing
processes. In a sense we have thus
identified a ``privileged'' distribution, since it can persist long both in
dynamic and in static (or quasi-static) equilibrium.

It may well be that by an appropriate choice of the initial spatial
distribution, the GAU1 initial mass function (or any other possible mass
function) would also stay practically unchanged, but this would be not so
interesting since, most likely, the required spatial distribution would be less
realistic. A systematic investigation of possible options in relation to
this issue is currently in progress and the results will be discussed in a
following article (Vesperini 1996, in preparation).

As for the difference between the mass function of inner and outer clusters,
in all of three simulations 
evolutionary processes tend to give rise to a  trend for inner
clusters to 
be on the average more massive than outer clusters (see van den Bergh
1995 for the 
observational  trend in our Galaxy and Crampton
et al. 1985 for the analogous trend in M31), though the extent of the
difference depends on the initial mass function (see Vesperini 1994
and Vesperini 1996, in preparation, for further details on this issue).

\subsubsection{ Spatial distribution and trends with galactocentric distance}
All runs start with a
spatial distribution such that the number density of clusters $n(R_g) \propto
R_g^{-3.5}$. Figure 12 shows the histogram of radial position at the beginning
and at the end of the simulation for the POW run (the histograms resulting from
the runs  GAU1 and GAU2 are similar). In all cases  almost all globular clusters
in the outer regions ($R_g>10 $ kpc) survive and the shape of the number density
distribution is essentially unchanged. In the GAU1 run, for which the mass
function is  'weaker', a slight depletion takes place also at larger
galactocentric distances.

Figures 13a-c show the initial and final CDF of
galactocentric distances for  all the simulations.  The CDF of observed
galactocentric distances of galactic globulars with $R_g<20$ kpc and $c \lta
1.7$ (as in our sample) is also shown. Evolutionary processes deplete the
central regions and modify the initial CDF making it very similar to the
observed one.

As for trends of structural parameters with the position inside the host
galaxy, it is interesting to note the result shown in Figure 14 where the
fraction of high concentration clusters (conventionally taken as those clusters
with $c>1$) is plotted as a function of the galactocentric distance  at the
beginning and  at the end of the simulation POW (analogous behaviour
was obtained for the runs GAU1 and GAU2). It is evident that evolution tends to
create a trend for more concentrated clusters to dominate near the galactic
center,  consistent  with the observational evidence (Chernoff \& Djorgovski
1989) that PCC clusters are preferentially located at small galactocentric
distances (see  Bellazzini
et al. 1996 for further comments on the trends between core parameters and
galactocentric distance).

\subsubsection{Rates of core collapse and evaporation}
As discussed in sect. 3 the method adopted for our investigation can not follow the evolution
of clusters after
they have reached the critical concentration for the onset of gravothermal
instability and so the code does not provide information about their
final fate; in particular, there remains the possibility that a few of these
objects  completely  evaporate before one Hubble time. Nevertheless,  it is
interesting to give the following  estimate of the fraction of the initial
number of clusters undergone core collapse and evaporation and of the current 
rates of core collapse and evaporation of globular clusters. We introduce two
quantities $\xi_D$ and $\xi_{CC}$, defined as the ratio of the number of clusters
undergoing  respectively  disruption and  core collapse before one Hubble time
to the total initial number of clusters. Another quantity of  interest is the
ratio of the number of core collapsed clusters to the total number of clusters
(core collapsed clusters plus normal 'King model' clusters) at the present epoch,
$f_{CC}$.

  In order to take into account the possibility of complete evaporation for 
clusters undergoing core collapse,  we have
counted among the disrupted clusters described by the parameter $\xi_D$ also
those undergoing core collapse if their evaporation timescale,
$t_{ev}=M/(dM/dt)$, calculated at the time $t_{cc}$ of the onset of
gravothermal 
instability, is sufficiently short, i.e. if $(t_{ev}+t_{cc})$ is smaller than
one Hubble time.
The actual value of the evaporation time is likely to be smaller than
$t_{ev}$ defined above (see, e.g., Lee \& Goodman 1995 for a detailed analysis
of the evaporation rate of PCC clusters) and  therefore, the values we supply
represent actually an upper limit for $\xi_{CC}$ and $f_{CC}$  and a lower limit
for $\xi_{D}$.

In our theoretical analysis we obtain for the GAU1 run  $\xi_{D}=0.610$,
$\xi_{CC}=0.145$, $f_{CC}=0.370$, for the GAU2 run  $\xi_{D}=0.418$,
$\xi_{CC}=0.188$, $f_{CC}=0.323$ and for the POW run
$\xi_{D}=0.587$, $\xi_{CC}=0.121$, $f_{CC}=0.292$. In our galaxy the observed
value of the frequency of core collapse clusters is $f_{CC}
\simeq 0.2$ (see, e.g., Trager, Djorgovski \& King 1993), in reasonable
agreement  with the above theoretical values. 

By the above calculation it is also possible to estimate the
current rate of clusters undergoing core collapse, $F_{cc}$, and the current
rate of cluster destruction, $F_{D}$ providing the fraction of the present
population of globular clusters undergoing core collapse or destruction per
Gyr:  $F_{cc}=0.05$ and $F_{D}=0.03$ for GAU1,  $F_{cc}=0.03$ and $F_{D}=0.03$
for GAU2 and  $F_{cc}=0.02$ and $F_{D}=0.05$ for POW.
The above values are in good agreement with those obtained
 by Hut \& Djorgovski (1992) on the basis of an analysis of the distributions of
the central and half-mass relaxation times of galactic globular clusters and of a
very simple  model for  their evolution; they 
estimate  $2\pm 1$ clusters per Gyr to be the current rate of core collapse and 
$5\pm 3$ clusters per Gyr  the current destruction rate, corresponding to a
fraction of the current number of galactic clusters of $F_{cc}\simeq 0.015\pm
0.007$ for the core collapse rate and of $F_{D} \simeq 0.038\pm 0.02$  for the
destruction rate. 

Finally we point out that our
simulations show that both $F_{cc}$ and $F_{D}$ may undergo a significant time
evolution and the extrapolation of the current values back in time might lead
to a wrong estimate of the past population of globular clusters. 
\section{Summary and conclusions}

After  setting the empirical background for our theoretical analysis with a
brief summary of the observational properties of the Galactic globular cluster
system, we have studied the evolution of globular clusters by means of a simple
and flexible semi-analytical model.

A preliminary study of the efficiency of various evolutionary
processes has been carried out by investigating the characteristic
time scales and following the evolution of individual globular
clusters starting from a large set of different initial conditions
without making any particular assumption on the initial conditions of the
galactic globular cluster system. This has allowed us to draw some
general conclusions on the origin (primordial or induced by evolution)
of a few interesting observational properties of the globular cluster
system of our Galaxy. These conclusions have received further support
by the results of a set of simulations following the evolution of
globular cluster systems starting from specific choices of initial
conditions for the initial mass function of the cluster system, for
its spatial distribution and for the initial distribution of concentrations.
By these simulations we have investigated how the relationships
between structural parameters may change toward the observational
results and in 
which way  the initial mass function and spatial distribution are modified as a
result of the evolutionary process. Although the main goal of 
this work was not a thorough comparison  with the
observations, nevertheless in some cases (spatial
distribution, mass function) our
results show   excellent consistency with the observational data.

 The main
results of our  analysis are the following:  
\begin{enumerate} 
\item Correlation between structural parameters. 

Evolutionary processes are unlikely to be responsible for 
the observed correlation between concentration and mass. In
fact, if low-concentration high-mass clusters were present at
the  time of cluster formation, many of them should still populate that region 
of the $c-\log M$ plane. Some high-concentration low-mass clusters are also 
expected to be present if concentration and mass were initially uncorrelated.

Our results point to a primordial origin for this correlation
with evolutionary processes giving rise to the
observed scatter. 

\item Mass function

Three simulations starting with different initial mass functions have been
carried out: a power-law initial mass function (POW),  a log-normal
with a mean 
value significantly smaller than the  observed value (GAU1) and a log-normal  
equal to the present mass function of galactic globular clusters (GAU2). 

In the
POW run the final mass function has a bell shape in the
logarithm of the mass resembling a Gaussian; if binned in $M$ the final mass
function can be described by a two-component power-law distribution.

In both the GAU runs
the gaussian shape of the mass function is preserved in the final sample. For
the GAU1 run, the mean value of $\log M$ undergoes a significant evolution from
the initial value $\langle \log M\rangle=4.5$ to a final value, $\langle \log
M\rangle=5.05$, approximately equal to the observed one.  Inspection
of the time 
evolution of $\langle \log M\rangle$  for the GAU1  run shows that after an
initial stage during which a significant change
 is observed, evolution slows down reaching a state of quasi-equilibrium. In
this 'quasi-equilibrium' state, although a significant disruption and core
collapse of clusters still take place, the value of $\langle \log M\rangle$ is
approximately constant. GAU2 run starts with an initial condition close to this
state of quasi-equilibrium and   both the shape and the parameters of the
initial mass function do not change during the evolution even though $70 \%$ of
the  clusters of the initial sample undergo disruption or core collapse before
one Hubble time and then are not included in the final sample.
\item Trends with galactocentric distance.

The observed trend for  more concentrated clusters to be located in the inner
regions has been shown to be a result of  evolution.

A trend for inner clusters  to be on average more massive than
those located in the outer regions is produced by evolutionary processes. The
extent of this effect depends on the initial mass function of the globular
cluster system and it is more pronounced if the initial mass function has a
larger fraction of low-mass clusters.

\item Spatial distribution.

A strong depletion of clusters in the very inner regions
of the galaxy has been observed. A good agreement between the
spatial distribution of clusters resulting from  our
simulations  and the observational data has been obtained in all the runs
done starting with $c$ and $\log M$ initially correlated.  
\end{enumerate}

Future work will extend this study by investigating the
evolution of globular cluster systems starting from a much larger
number of different initial conditions (Vesperini 1996, in preparation).
Moreover we plan to study the dependence of the above
results on the model for the host galaxy  (see Vesperini 1994 for some
preliminary results) and to include processes, such as  bulge-shocking,
that are expected to occur in the presence of non-circular orbits. 
\section*{Acknowledgments}
I wish to thank  Giuseppe Bertin  for many enlightening discussions. I am very
grateful to him  for his very careful reading of this paper and for the numerous
suggestions and comments.\\
I wish to thank  T.S. van Albada, D.C. Heggie and M.
Stiavelli for many useful discussions.$~~~~~~~~~$
\section*{References}
Abraham R.G., van den Bergh S., 1995, ApJ, 438, 218\\  
Aguilar L., Hut P., Ostriker J.P., 1988, ApJ, 335, 720\\ 
Bahcall J.N., 1984, ApJ, 287, 926\\  
Bahcall J.N., Schmidt M., Soneira R.M., 1982, ApJL, 258, L23\\
Bellazzini M., Vesperini E., Ferraro F.R., Fusi Pecci F., 1996, MNRAS,
279,337\\ 
Binggeli B., Sandage A., Tarenghi M., 1984, AJ, 89, 64\\ 
Binney J., Tremaine S. 1987, Galactic Dynamics, 

 Princeton University Press, Princeton, New Jersey\\
Caputo F., Castellani V., 1984, MNRAS, 207,185\\
Chernoff D.F., Kochanek C.S., Shapiro S.L., 1986, ApJ, 309, 183 [CKS]\\ 
Chernoff D.F., Shapiro S.L., 1987, ApJ, 322, 113\\ 
Chernoff D.F., Djorgovski S.G., 1989, ApJ, 339, 904\\ 
Crampton D., Cowley A.P., Schade D., Chayer P., 1985, ApJ, 288, 494 \\
Djorgovski S.G., 1991, in 'Formation and Evolution of Star
Clusters',  

ed. K.Janes, ASP, p. 112 \\
Djorgovski S.G., Meylan G., 1994, AJ, 108,1292 [DM]\\ 
Fall S.M., Rees M.J., 1977, MNRAS, 181,27p\\
Fall S.M., Rees M.J., 1985, ApJ, 298, 18\\
Gnedin O.Y., Ostriker J.P., 1997, ApJ, 474, 223\\
Harris W.E., 1991, ARAA, 29, 543\\
Harris W.E., Pudritz R.E., 1994, ApJ, 429, 177\\
Heggie D.C., Aarseth S.J., 1992, MNRAS, 257, 513\\
Hut P., Djorgovski S.G., 1992, Nature, 359, 806\\
Ichikawa S., Wakamatsu K., Okamura S., 1986, ApJS, 60, 475\\
Katz J., 1980, MNRAS, 190, 497\\
King I. R., 1965, AJ, 70, 376\\
King I. R., 1966, AJ, 71, 64\\ 
Lauer T. R., 1985, ApJ, 292, 104\\
Lee H.M., Goodman J., 1995, ApJ, 443, 109\\
Mandushev G., Spassova N., Staneva A., 1991, A\&A, 252, 94\\
McLaughlin D.E., 1994, PASP, 106, 47\\
Murali C., Weinberg M., 1996, MNRAS, submitted\\
Prata S.W., 1971a, AJ, 76, 1017\\
Prata S.W., 1971b, AJ, 76, 1029\\
Secker J., 1992, AJ, 104, 1472 \\
Spitzer L., 1987, 'Dynamical Evolution of Globular Clusters', Princeton 

University Press\\
Trager S.C., Djorgovski S.G., King I., 1993 in : Structure and Dynamics 

of 
Globular Clusters, p.343, eds. S.G. Djorgovski \& G. Meylan ASP\\
van den Bergh S., 1993, in : Structure and Dynamics 

of 
Globular Clusters, p.1, eds. S.G. Djorgovski \& G. Meylan ASP\\ 
van den Bergh S., 1994, ApJ, 435, 203\\
van den Bergh S., 1995, AJ, 110, 1171\\
Vesperini E., 1994, Ph.D. Thesis, Scuola Normale Superiore, Pisa\\
Vesperini E., Chernoff D.F., 1994, ApJ, 431, 231\\
Vietri M., Pesce E., 1995, ApJ, 442, 618\\
Weinberg M., 1994a, AJ, 108, 1414\\
Weinberg M., 1994b, AJ, 108, 1403\\
Wiyanto P., Kato S., Inagaki S., 1985, PASJ, 37, 715\\
Zinn R., 1985, ApJ, 293, 424\\
\newpage
\clearpage
\section*{Figure captions}
Figure 1  Evaporation (straight lines) and disk shocking timescale versus
galactocentric distance. Average of timescales corresponding to $W_0=1,~3,~4,~7$
have been plotted.\\
Figure 2 Survival factor versus galactocentric distance for different initial
values of the cluster mass.\\
Figure 3 Initial conditions in the plane $W_0-R_g$ of clusters surviving
disruption and core collapse. The evolution of clusters with initial conditions
in the grid $2<W_0<7$, $3<R_g(\hbox{kpc})<18$  has been followed.
The points plotted are those initial conditions that do not lead to the
disruption or core collapse of clusters in one Hubble time. Empty regions
indicate initial conditions for which the cluster undergoes disruption or core
collapse before one Hubble time. Different symbols indicate different
values of the initial mass: squares  $M=3\times
10^4 M_{\odot}$, crosses  $M=3\times 10^5 M_{\odot}$, circles 
$M=3 \times 10^6 M_{\odot}$.\\ 
Figure 4 Final $W_0$ versus galactocentric distance
for clusters surviving disruption and core collapse whose initial conditions are
shown in Figure 3. Symbols are as in Figure 3.\\ 
Figure 5 Concentration versus logarithm of the mass of clusters at the end of
the POW run; the solid line shows the initial relationship between
$c$ and $\log~M$ in the simulation.\\
Figure 6 Concentration versus logarithm of the mass of clusters at the end of
the GAU1-nc run.\\
Figure 7 Histogram of the initial (grey) and final (stripes) values of masses of
clusters for the POW run (a), the GAU1 run (b) and the GAU2 run (c).\\
Figure 8 Initial (short dashed line), final (solid line) and observational (long
dashed line) cumulative distribution function of the mass of clusters for the POW
run (a), the GAU1 run (b) and the GAU2 run (c).\\
Figure 9 Initial (open squares) and final (full squares) mass function in
$M$ from the POW run; lines show the best fit.\\
Figure 10 (a) Time evolution of the mean value of the logarithm of the
mass of clusters for the GAU1 run (full dots) 
and for the GAU2 run (full triangles); (b) Time evolution of the
number of clusters in the simulations for the GAU1 run (full dots) and
for the GAU2 run (full triangles).\\ 
Figure 11 Cumulative
distribution function of the logarithm of the mass of clusters in the GAU1 run
at different times: from the left to the right the curves corresponds to t=0 Gyr
(thin solid line), t=2 Gyr (dot-dashed line), t=4 Gyr (dotted line), t=9 Gyr
(dashed line), t=15 Gyr (thick solid line).\\
Figure 12 Histogram of the galactocentric distances for the initial (grey) and
final (stripes) sample of clusters from the POW run. No cluster with
$R_g>8$ Kpc undergoes core collapse or disruption and thus the initial
and the final histograms are exactly superimposed.\\
Figure 13 Initial (short dashed line), final (solid line) and observational
(long dashed line) cumulative distribution function of galactocentric
distances for the POW run (a), the GAU1 run (b) and the GAU2 run (c).\\
Figure 14 Ratio of the number of clusters with $c>1$ to the total number in
each radial bin versus galactocentric distance for the POW run. (open squares
indicate initial values, full squares indicate final values).\\
 \end{document}